\documentclass[12pt]{article}

\usepackage{graphicx}
\usepackage{subfigure} 

\usepackage[left=2.5cm,right=2.5cm,top=2cm,bottom=2cm]{geometry}

\usepackage[UKenglish]{babel} 
\usepackage{comment}
\usepackage{multicol}
\usepackage{array} 
\usepackage{amssymb} 
\usepackage{amsmath}
\usepackage{xspace} 
\usepackage{amsbsy} 
\usepackage{layout}

\newcommand{\celsius}{\ensuremath{\mathrm{^oC}}\xspace} 
\newcommand{\gammaprime}{\ensuremath{\gamma^{\prime}}\xspace}

\newcommand{\gammachem}{\ensuremath{\mathrm{AlAg_2}}\xspace}

\newcommand{\planes}[2][\alpha]{\ensuremath{\mathrm{\{#2\}_{#1}}}\xspace} 

\newcommand{\fig}[2][0.45]{\resizebox{#1\textwidth}{!}{\includegraphics{#2}}} 
\newcommand{\burger}[3]{\ensuremath{\boldsymbol{b}=\frac{#1}{#2} [ #3 ]  }\xspace} 

\begin{document}

\title{Nucleation and growth of the $\gammaprime (\gammachem)$ precipitate in Al-Ag(-Cu) alloys}
\author{Julian~M.~Rosalie$^{a,b,c}$ 
\and Laure~Bourgeois$^{b,c,d}$ \and
Barrington~C.~Muddle$^{b,c}$}
\date{}

\maketitle
\noindent
$^a$Microsctructure Design Group, Structural Metals Center, National Institute for Materials Science (NIMS), Japan.\\
$^b$ARC Centre of Excellence for Design in Light Metals, Australia.\\
$^c$Department of Materials Engineering, Monash University, 3800, Victoria, Australia.\\
$^d$Monash Centre for Electron Microscopy, Monash University, 3800, Victoria, Australia.

\begin{abstract}
Precipitation of the \gammaprime (\gammachem) phase was investigated in Al-Ag(-Cu) alloys using 
high resolution transmission electron microscopy and scanning transmission electron microscopy.
Precipitation commenced with segregation of silver to stacking faults, 
followed by thickening in steps corresponding to single unit cell ledges. 
In conjunction with gradual segregation of silver and aluminium into ordered layers this yielded \gammaprime  phase platelets with a thickness of either 
$2 \mbox{ or } 3\times$ the \gammachem $c$-lattice parameter
Plates with a thickness of $2 c(\gammachem)$ would not achieve self-accommodation of the shape strain for transformation. 
Further thickening of the precipitates was slow, despite considerable silver segregation around the precipitates. 
Growth by the addition of single unit cell height ledges is expected to lead to an additional shear strain energy barrier to ledge nucleation and this may contribute to a process of nucleation-limited, growth.
\end{abstract}



\section{Introduction}

Aluminium-silver alloys provide a valuable model system for understanding precipitate nucleation and growth in aluminium alloys. 
Precipitation of the \gammaprime (\gammachem) phase involves a structurally simple face-centred cubic (fcc) $\rightarrow$  hexagonal close-packed (hcp) transformation.  
The transformation is exceptional since it involves minimal volumetric strain \cite{nie:1999}.
Clarifying precipitation in such systems is required to understand and enhance precipitation in more complex aluminium alloys such as those prevalent in aerospace and other transport applications.

The \gammaprime (\gammachem) phase has a layered structure with alternate high and low silver layers  \cite{howe:1987,zarkevich:2002} whose composition
 can depart considerably from stoichiometry \cite{moore:2000}.
There is exceptionally good 1:2 matching between the interplanar spacing of Al \planes[]{111} planes (0.23\,nm) 
and the \gammaprime basal plane ($c$ = 0.4607\,nm \cite{howe:1987}).
This permits the formation of high-aspect ratio plates in Al-Ag  \cite{zakharova:1966,nicholson:1961, howe:1987a} and Al-Ag-Cu alloys \cite{bouvy:1965, khatanova:1966, borchers:1969a}, in which  the basal plane retains coherence with the $\{111\}_\mathrm{Al}$ habit plane. 

Thickening of the \gammaprime  phase is thought to occur  by the progression of kinked ledges across the habit plane \cite{aikin:1987, aikin:1991, csontos:1998} as is common for plate-shaped precipitates \cite{shiflet:1998}. 
The passage of a ledge with Shockley partial dislocation (Burgers vector, \(\boldsymbol{b}=\frac{a}{6}\langle112\rangle\))  on every second close-packed plane 
would accomplish the structural change for the phase transformation \cite{aikin:1990}, changing the stacking sequence from $ABCA (fcc) \rightarrow ABA$ (hcp). 

Single ledges are thought to coalesce into pairs rather than acting individually \cite{aikin:1990} and growth ledges with heights of  2, 4 and 6 
$\times c_{AlAg2} $ have been measured \cite{howe:1985a}. 
The coalescence of two  Shockley dislocations would yield a perfect dislocation and ledges corresponding to both \burger{a}{2}{110} and $\frac{a}{2}[1\overline{1}0]$ have been observed in  equal numbers \cite{howe:1985a} on the interfaces of large \gammaprime precipitates \cite{howe:1985}.
It was thus inferred that growth involved all three variants of the Shockley partial dislocation  in the habit plane in equal numbers to balance the transformation-induced strain \cite{howe:1985}.  

It has been proposed that the nucleation  of a hcp phase in a fcc matrix could be accomplished homogeneously in an analogous manner  
through the  formation of a ``self-accommodating'' unit in which transformation dislocations in each layer were balanced by those in  adjacent layers \cite{shchegoleva:1976}. 
This requires a minimum thickness of seven atomic close-packed layers 
($3\times c$(\gammachem)) to allow the outer-most atomic planes of the precipitate to maintain coherence with the matrix.

There is good evidence, however, that precipitation of the \gammaprime precipitate is heterogeneous and closely associated with stacking faults. 
Precipitation has been observed on Frank dislocation loops \cite{frank:1961,nicholson:1961} and dislocations  which provide a faulted surface \cite{nicholson:1961,frank:1961,  passoja:1971}.
When the \gammaprime phase formed on dislocation loops,  precipitation occurred only on the faulted surfaces \cite{Rosalie2009a,Rosalie2009b}. 
In addition, silver-rich GP zones that form in advance of the \gammaprime phase persist in the microstructure despite prolonged ageing \cite{nicholson:1961}, indicating poor nucleation behaviour.

The phase transformation involves a change in the composition of the product phase in addition to the structural rearrangement.
For coherent plate-like precipitates this compositional adjustment occurs by diffusion across the disordered riser of the ledges. 
These structural and compositional components of the phase transformation are interconnected since the local composition affects the stacking fault energy (SFE) and therefore the preferred structure. 
Pure Al is stable against stacking fault formation \cite{finkenstadt:2006} with estimates of the SFE  ranging from 117-130 $\mathrm{mJ.m}^{-2}$  \cite{laird1969,ramanujan:1992} to  $200\mathrm{mJ.m}^{-2}$  \cite{gallagher:1970}. 
However, the SFE of Ag is only 22\,$\mathrm{mJ.m}^{-2}$ \cite{gallagher:1970} and 
 calculations show that Ag lowers the local SFE in sold solution such that for 50--90 at.\% Ag  a hcp structure is preferred \cite{schulthess:1998} .

Calculations using the Vienna \textit{ab initio} simulation package ({\scshape vasp})  led to the proposal that precipitation of the \gammaprime phase commences with segregation of silver to stacking faults, forming a structure initially comprised of pure Al and Ag layers  \cite{finkenstadt:2006}.
Ag segregation to stacking faults was energetically favourable \cite{finkenstadt:2006} due to the lower stacking fault energy (SFE) for Ag and Ag-rich regions of the matrix \cite{schulthess:1998}.
The local hcp structure at the stacking fault would ensure that a three close-packed layer structure ($1\times c(\gammachem)$) could be formed via diffusion with no additional structural rearrangement. 
The existence of such structures has not been proven, and there is a lack of understanding as to how the precipitate might progress to a strain-accommodated form. 

Previous experimental work on the thickening of the \gammaprime precipitates has been conducted on precipitates of much larger scale \cite{howe:1985,howe:1985a,Finkenstadt2009}, which may not accurately reflect the situation during nucleation. 
The present work involved a close examination of the \gammaprime phase in  Al-Ag(-Cu) alloys subjected to  short-term ageing treatments in order to better understand the initial stages of growth. 
This sought to clarify the  nucleation mechanism of the precipitate and in particular whether the precipitate initially adopts a self-accommodating form. 

\section{Experimental details}
The alloys used in this work was cast from high-purity aluminium (Cerac alloys, 99.99\% purity), 
silver (AMAC alloys, 99.9+\%) and copper (AMAC alloys, 99.99\%). 
The compositions used were Al-1.68at.\%Ag  and Al-0.9at.\%Ag-0.90at.\%Cu.

The pure metals were melted in air at 700\celsius in a graphite crucible, 
stirred and poured into graphite-coated steel molds. 
The cast ingots were homogenised at 525\celsius for 7\,days.
Ingots were later hot- and cold-rolled to 0.5\,mm for analysis.
Alloy compositions were determined by inductively-coupled plasma atomic emission spectrometry (ICP-AES). 
Solution treatments were carried out on 3\,mm diameter $\times$ 0.5\,mm thickness discs for 0.5\,h at 525\celsius in a nitrate/nitrite salt pot. 
After this point, the material was subjected to no further deformation. 
After quenching into water at ambient temperature, 
samples were aged at 200\celsius using oil baths for 0.5--4\,h.

Foils for electron microscope analysis were jet electropolished at \(-20\celsius\) using a nitric acid/methanol solution 
(33\% \(\mathrm{HNO_3}\) / 67\%\(\mathrm{CH_3OH}\) v/v). 
The applied voltage was $-13$\,V and the current averaged 200\,mA.

HRTEM examination made use of a JEOL~2011 HRTEM operating at 200\,kV with a point resolution of 0.23\,nm.
This provided accurate information on the lattice position and was used for identifying displacements of the matrix  planes occurring across the precipitates. 
However, image delocalisation at the precipitate-matrix interfaces made the precipitate thicknesses difficult to measure precisely. 

High angle annular dark-field (HAADF) STEM experiments and energy dispersive x-ray (EDX) mapping were conducted on a JEOL 2100F instrument, operating at 200\,kV with TEM point resolution of 0.23\,nm and STEM resolution of 0.19\,nm.
The convergence semi-angle used was 13\,mrad and the inner and outer collection angles were 65 and 185\,mrad respectively, resulting in images with a 1.9\AA  ~resolution and dominated by Z contrast.
This permitted clearer imaging of the precipitate-matrix interfaces, but was subject to scanning noise and distortion which sometimes made interpreting long-range matrix displacements problematic. 
(The STEM images shown are as collected without any image manipulation aside from from contrast and brightness adjustment, except where explicitly stated.)
Using both techniques in a complementary fashion allowed for analysis of both the \gammaprime precipitate thickness and the matrix deplacements due to the transformation dislocations. 

\section{Results}

\begin{figure*}
\begin{center}
\subfigure[Al-Ag (0.5\,h 200\celsius)  \label{fig-tem-1}]{\fig[0.48]{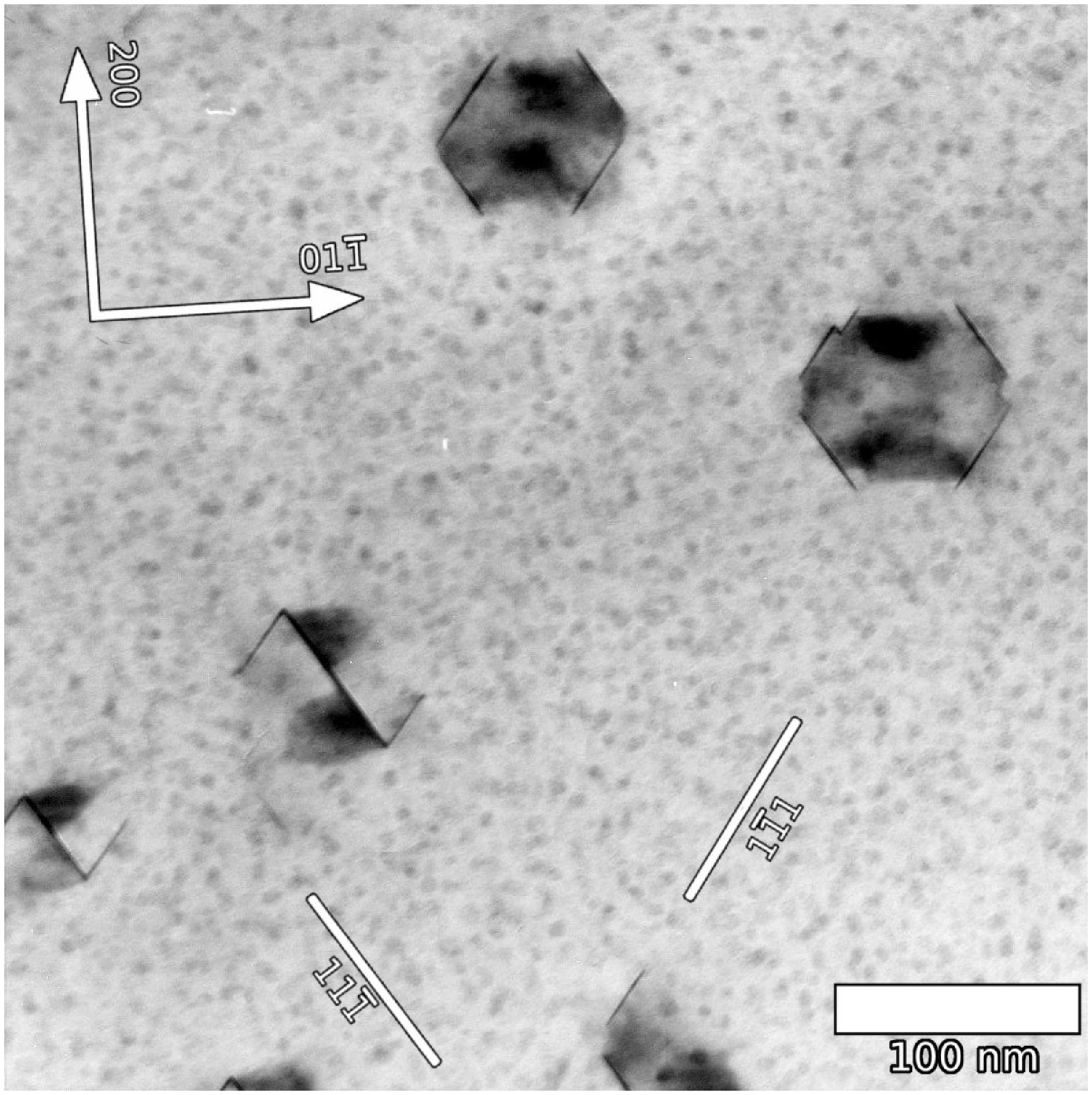}}
\subfigure[Al-Ag-Cu (0.16\,h, 200\celsius) \label{fig-tem-2}]{\fig[0.48]{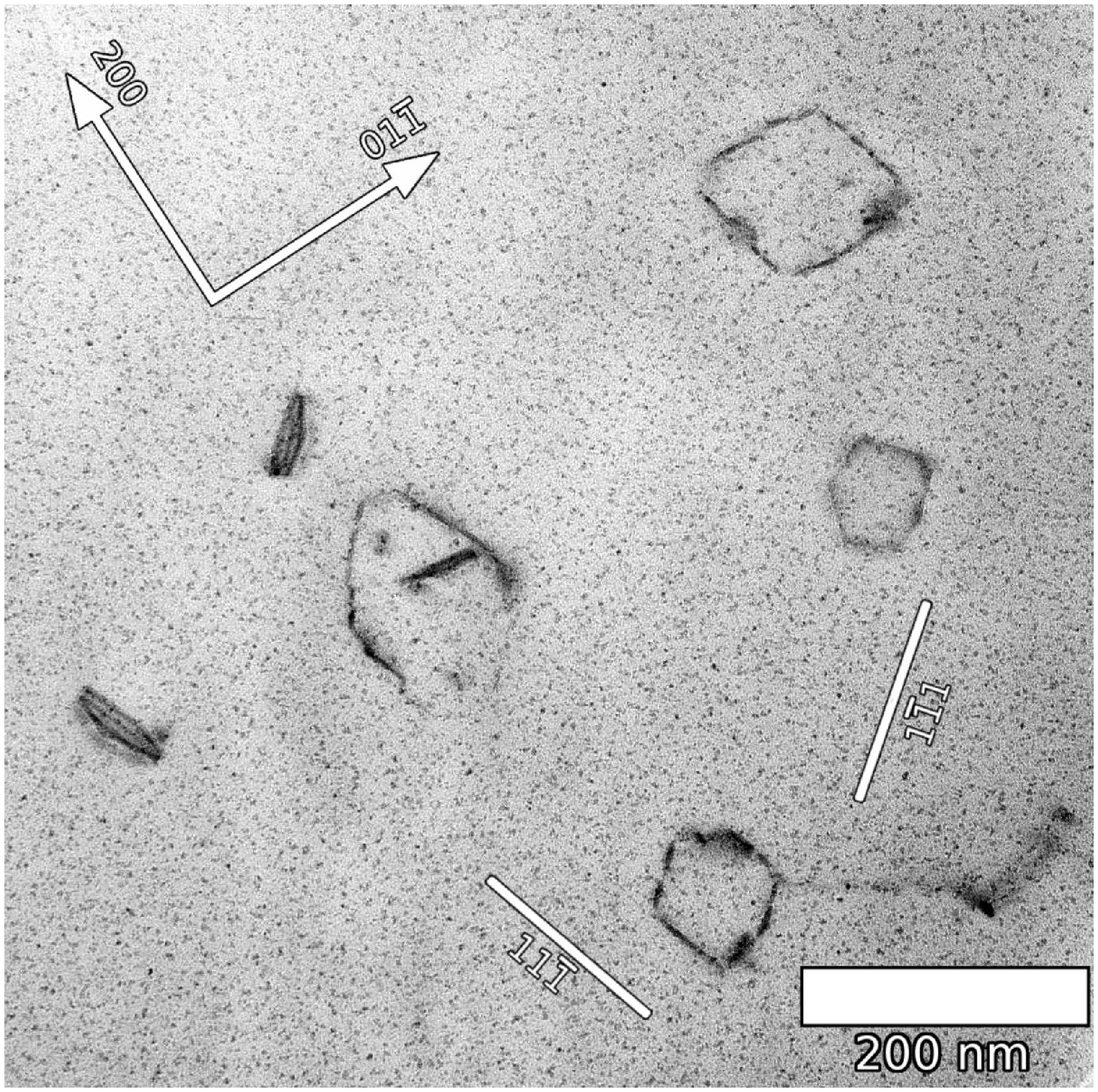}}
\caption{Bright field TEM images of the microstructures of (a) Al-Ag and (b) Al-Ag-Cu alloys.
The images were obtained with the electron beam along [011]$_\mathrm{Al}$.
Both alloys contained \gammaprime precipitates arranged in structured assemblies. 
\label{fig-tem}}
\end{center}
\end{figure*}

The Al-Ag (Figure~\ref{fig-tem-1}) and Al-Ag-Cu  (Figure~\ref{fig-tem-2})  alloys aged for $\le$0.5\,h at 200\celsius contained Ag-rich Guinier-Preston (GP) zones and a sparse distribution of \gammaprime precipitates.
The precipitates formed structured assemblies in association with dislocation loops that have been 
 described previously \cite{Rosalie2009a,Rosalie2009b} and are not discussed in detail here.

\subsection{HRTEM}

\begin{figure*}
\begin{center}
\subfigure[\label{fig-hrtem-a}]{\fig[0.48]{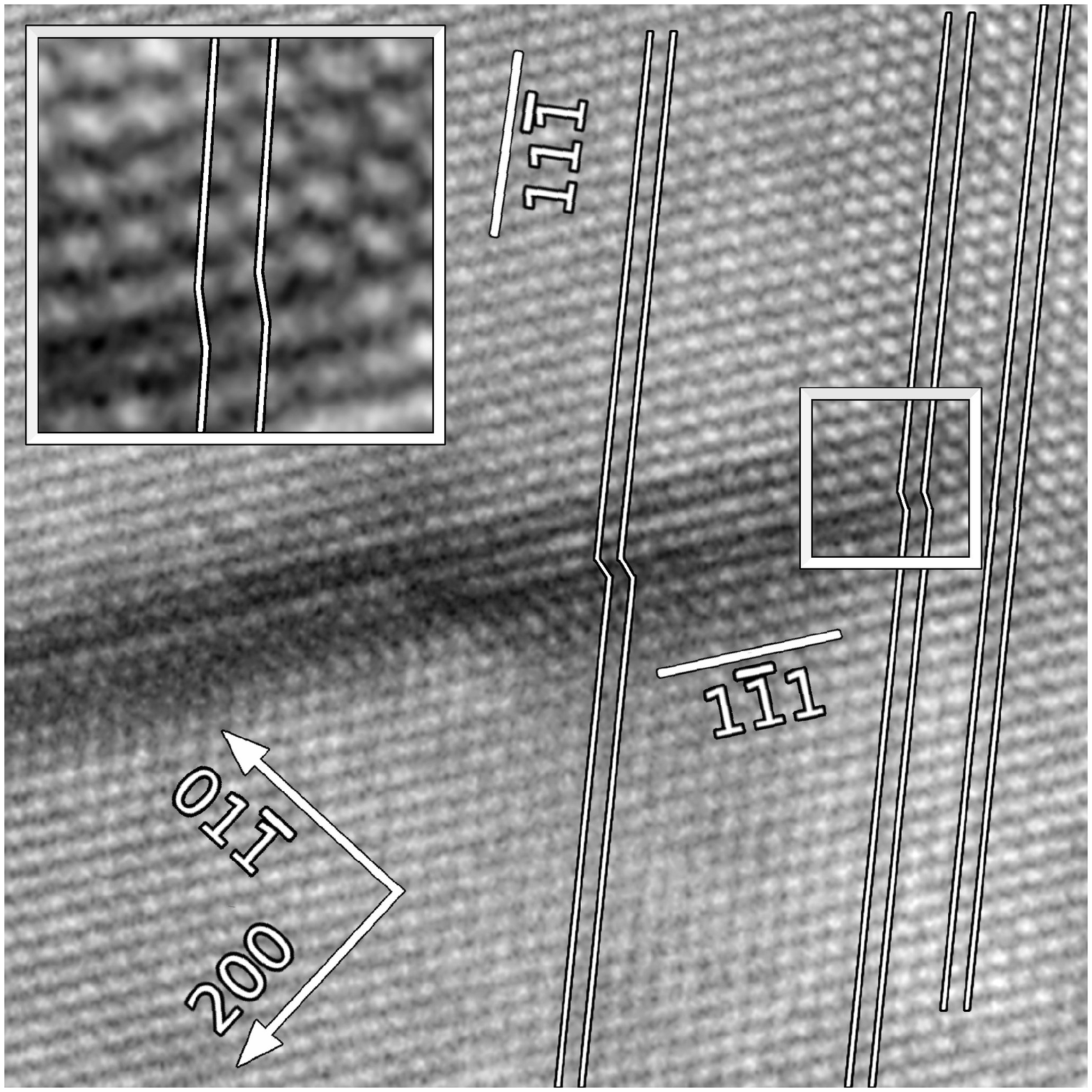}} \hfill
\subfigure[\label{fig-hrtem-b}]{\fig[0.48]{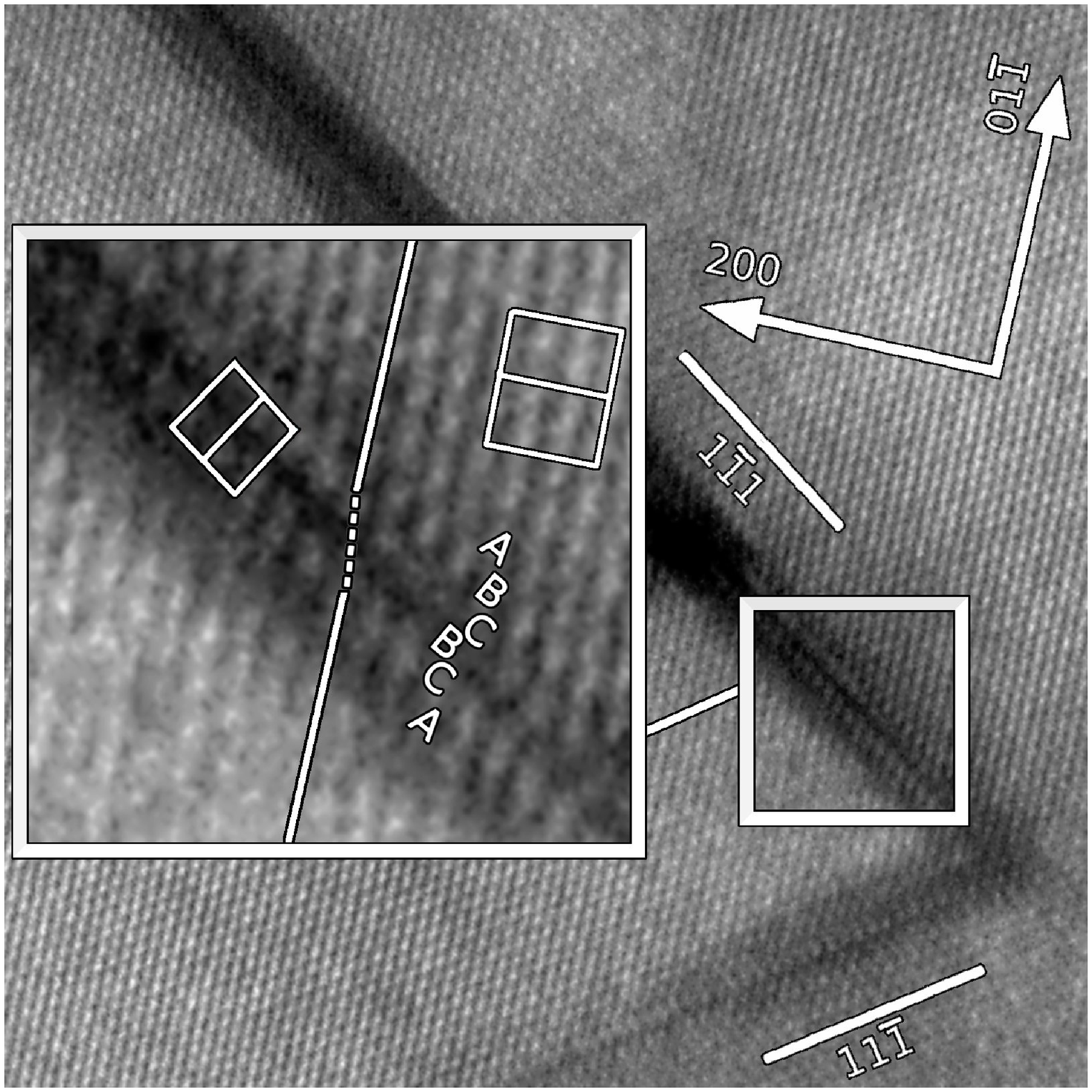}} 
\caption{A larger diameter \gammaprime precipitate.
The HRTEM image shows a precipitate in a foil of Al-Ag aged for 0.5\,h and 200\celsius. 
The contrast within the precipitate shows alternate bright-dark contrast in the (111) planes, suggesting ordering.
The matrix planes are displaced in passage across the precipitate. 
A single stacking fault can be seen, allowing for the presence of a single unit cell thickness with hcp structure. 
\label{fig-htrem}}
\end{center}
\end{figure*}

HRTEM images of the earliest resolvable precipitates in Al-Ag and Al-Ag-Cu showed structures consisting of several silver-enriched \{111\} planes surrounding a single stacking fault.
Figure~\ref{fig-hrtem-a} shows a precipitate formed after 0.5\,h ageing at 200\celsius.
Lines have been drawn along the local intensity maxima (bright spots) of the close-packed planes to highlight the presence of lateral displacements (which can be seen most clearly at a glancing angle). 
Lines drawn along the matrix show no deviation, indicating that the atomic planes remained in good alignment over considerable distances in essentially defect-free crystal.
However, lines drawn through the precipitate show that the matrix planes on either side of the precipitate are incommensurate. 
Close inspection  (see enlarged inset) shows that the displacement of the matrix planes has occurred along a single
 plane within the precipitate, indicating the presence of a single stacking fault.
In addition, diffraction TEM images showed no sign of misfit dislocations at the planar matrix-precipitate interfaces. 
This would indicate that the 
\( \{0001\}_{\gammaprime} \parallel \{111\}_{\alpha} \) interfaces were coherent (as expected for \gammaprime precipitates, particularly of such small size).

\subsection{STEM}

Silver-enriched regions were readily distinguishable from the matrix using HAADF STEM imaging due to the large difference in atomic number between Ag (atomic number, Z=47) and Al (Z=13).

\begin{figure*}
\begin{center}
\subfigure[Al-Ag\label{fig-haadf-alag}]{\fig[0.48]{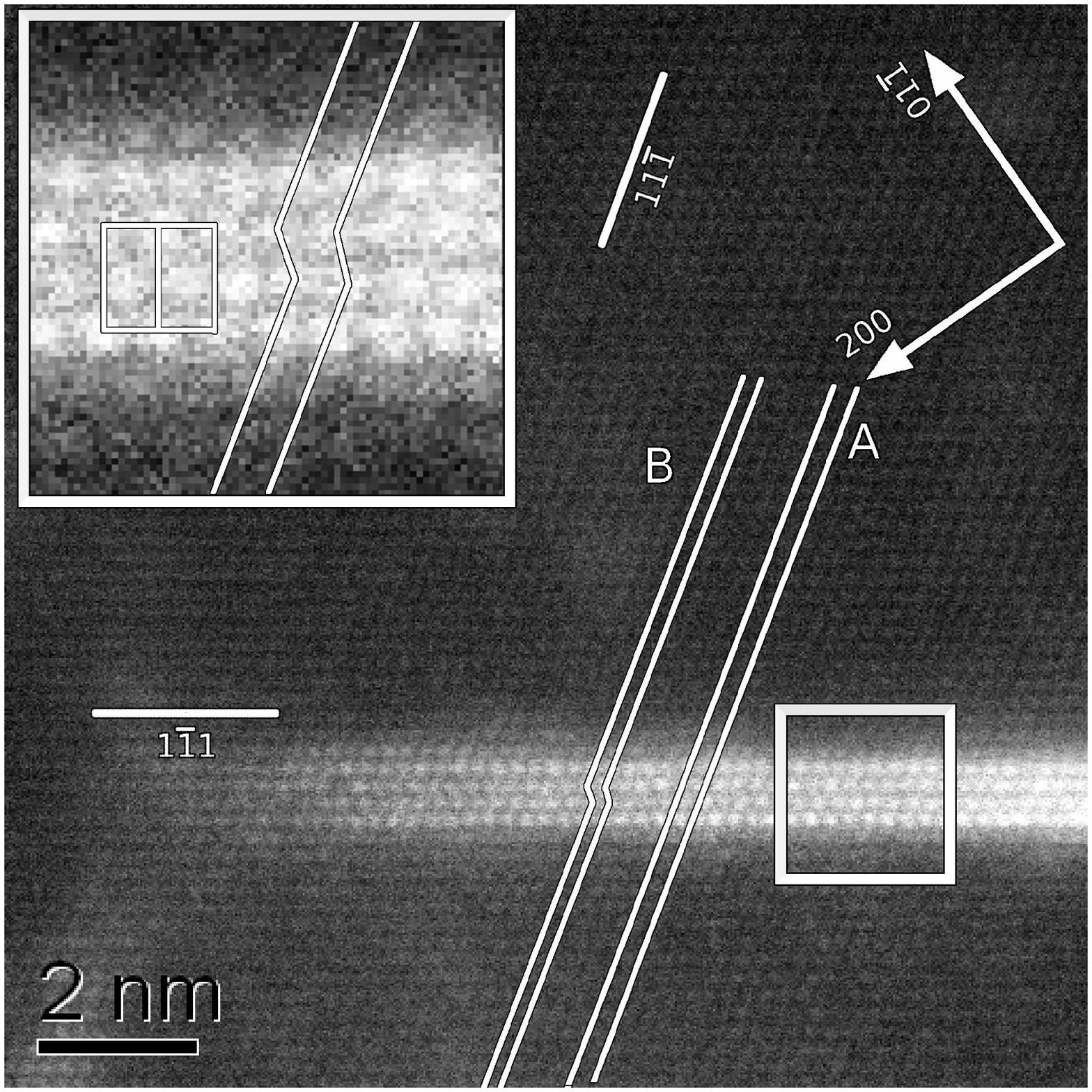}}\hfill 
\subfigure[Al-Ag-Cu\label{fig-haadf-alagcu}]{\fig[0.48]{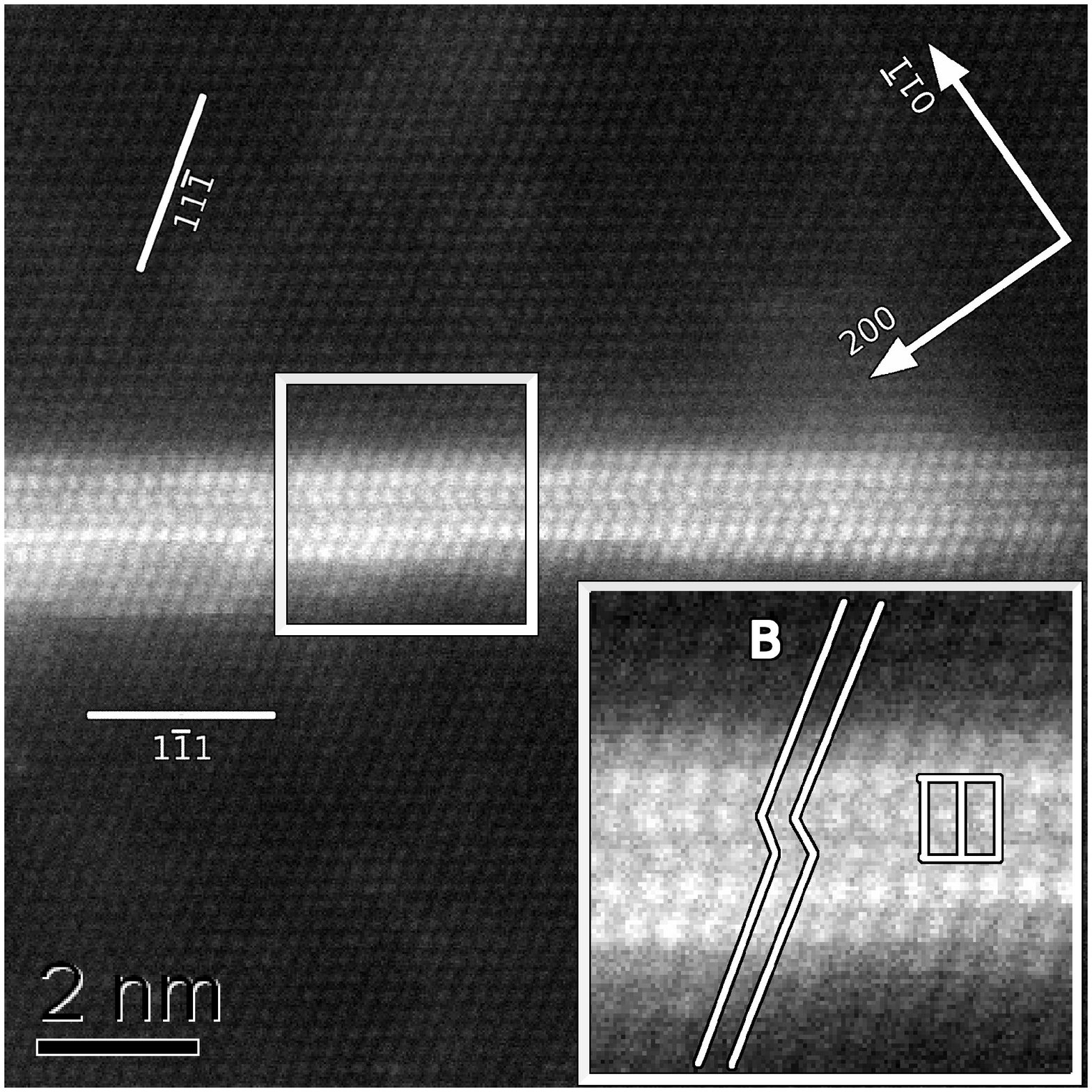}}\hfill 
\caption{HAADF STEM images from precipitates in (a) Al-Ag and (b) Al-Ag-Cu. 
Both foils were aged for 0.5\,h at 200\celsius. 
The insets shows an enlarged view of the precipitate.
In a) Lines (labelled A) along the $(11\overline{1})$ planes show that the matrix planes are incommensurate across the precipitate. 
A second series of lines (labelled B) show that this displacement takes place along a single close-packed $(1\overline{1}1)$plane.
A unit cell of the hexagonal close-packed structure is indicated.
\label{fig-stem-a}}
\end{center}
\end{figure*}

Figure~\ref{fig-stem-a} shows HAADF STEM images of diffracted intensity for (a) Al-Ag and (b)Al-Ag-Cu alloys aged for 0.5\,h. 
Lines drawn along the matrix $(11\overline{1})$ planes (labelled A) in Figure~\ref{fig-haadf-alag} are displaced in the habit plane of the precipitate and the matrix planes on either face of the precipitate are incommensurate. 
Stacking faults are clearly visible in the images, where lines  along the (111) planes (labelled ``B'') are displaced in the habit plane.
A single hcp unit cell is indicated in each enlarged inset. 
Similar displacements are also visible in Figure~\ref{fig-haadf-alagcu}. 
The diffracted intensity in the four central layers was slightly greater in the outermost layers but there was no clear evidence of alternating weak-strong atomic contrast within the precipitate.
Outside this central zone, the transmitted intensity decreased to that in the matrix over a distance of 2--3\,nm.  
To the left the outer enriched zone is considerably broader (6--8 layers) than it is to the right; however there is no further change in the lattice positions.   
Although it is not possible to directly distinguish whether the stronger contrast around the precipitates is due to Cu or Ag, there was no recognisable difference between precipitates in Al-Ag and Al-Ag-Cu under the present ageing conditions. 
In addition, EDX mapping in the Al-Ag-Cu alloy after ageing to 0.5\,h failed to show any segregation of copper to the precipitates or stacking faults, whereas silver was heavily enriched in these regions.  
Both factors suggest that Cu segregation to the defect and/or precipitate is negligible.

\begin{figure*}
\begin{center}
\subfigure[Bright field (BF) STEM\label{fig-bf}]{\fig[0.48]{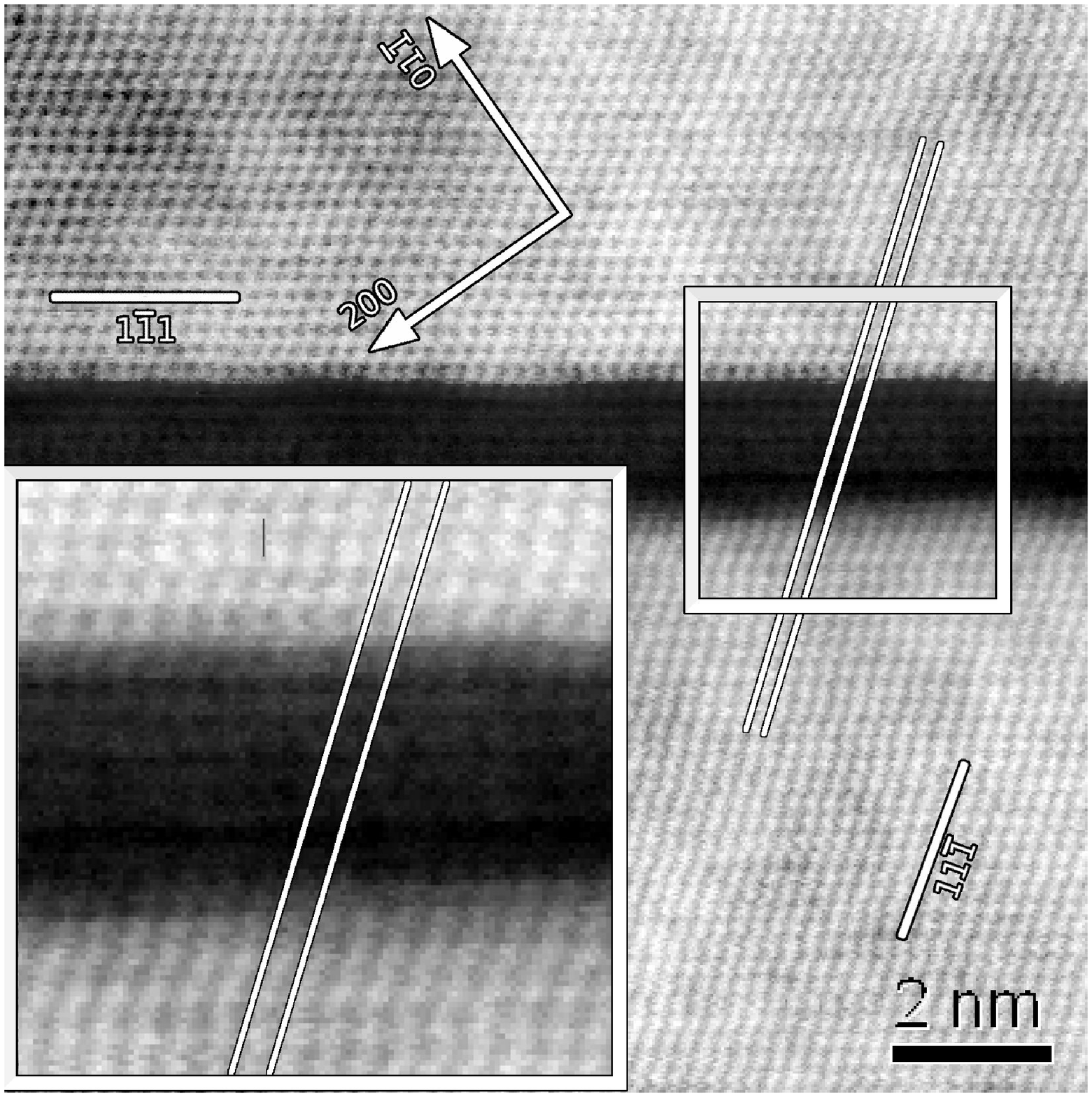}} \hfill 
\subfigure[HAADF STEM\label{fig-bf}]{\fig[0.48]{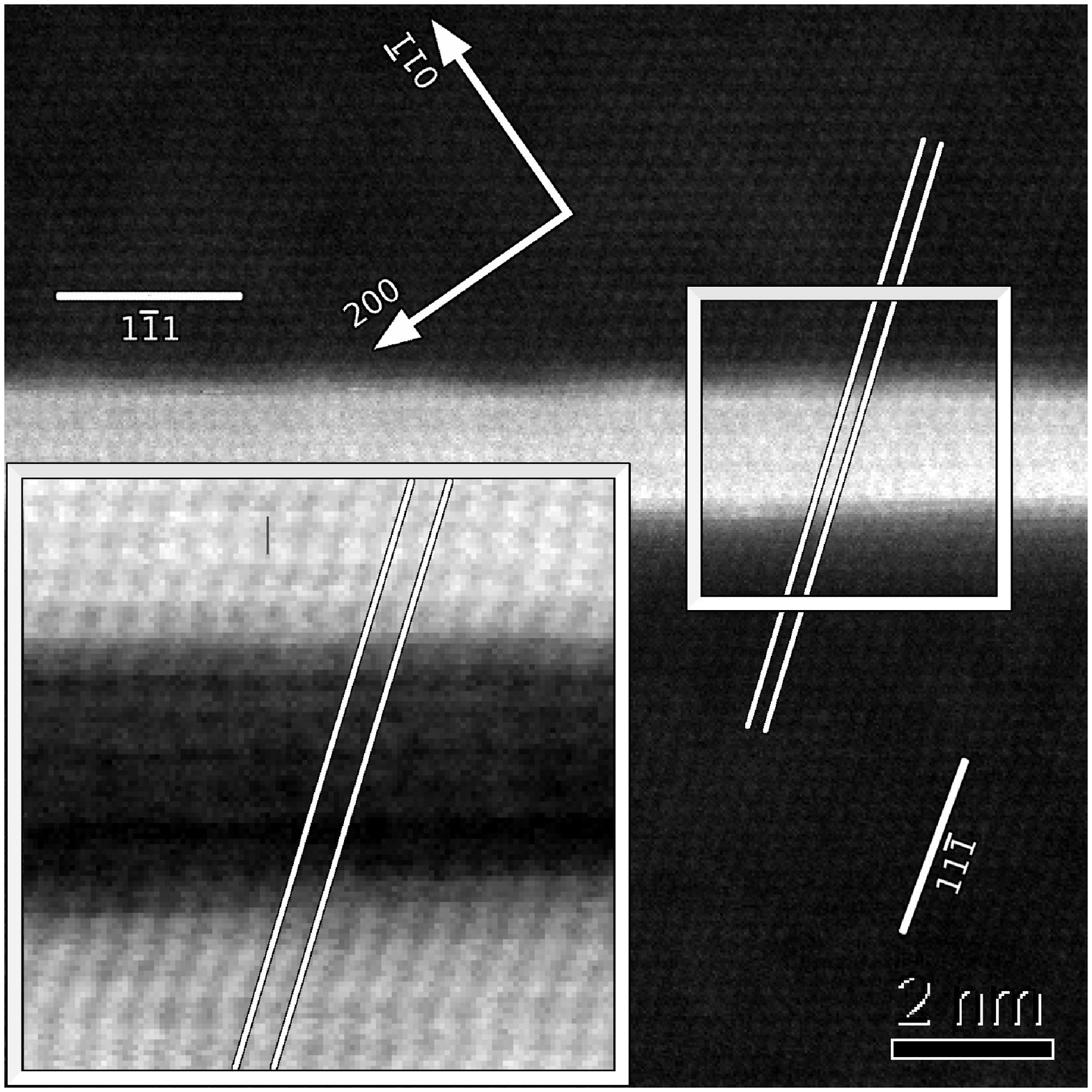}}
\caption{STEM images of a precipitate in Al-Ag-Cu aged for 4\,h at 200\celsius.
a) and b) show bright field (BF) and HAADF STEM images, respectively. 
The precipitate is of greater thickness than the example presented in  Figure \ref{fig-stem-a} and the precipitate lattice is less clearly resolved.
Lines drawn along $(11\overline{1})$ planes show that the matrix planes are incommensurate across the precipitate.
\label{fig-stem-4h}}
\end{center}
\end{figure*}

The local structure was less clear in thicker precipitates, such as the example presented in Figure \ref{fig-stem-4h} which shows (a)  bright field (BF) and  (b) HAADF STEM images from a sample of Al-Ag-Cu aged for 4\,h at 200\celsius. 
Slight drift can be observed in the image; however, lines drawn across the precipitate still indicate clearly that the matrix planes on either face are incommensurate despite the greater thickness.

\begin{figure*}
\begin{center}
\subfigure[Bright field (BF) STEM\label{fig-bf-layers}]{\fig[0.48]{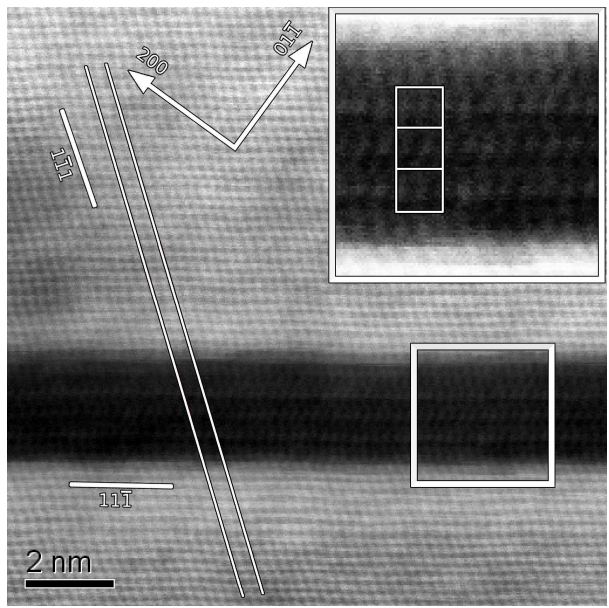}} \hfill 
\subfigure[HAADF STEM\label{fig-haadf-layers}]{\fig[0.48]{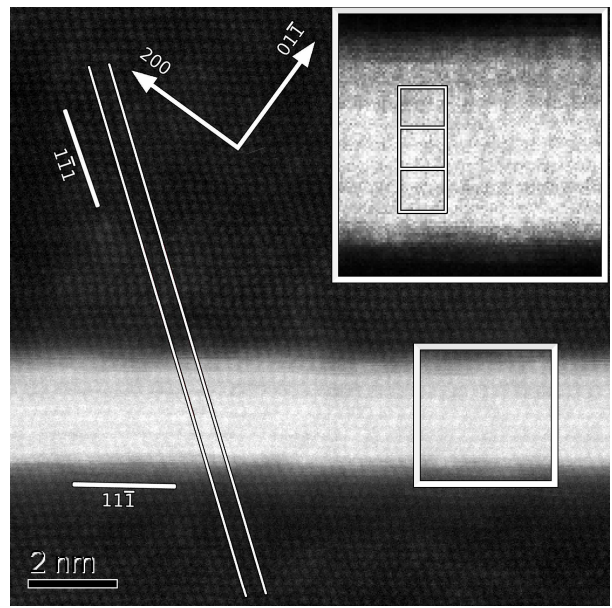}} 
\caption{STEM images of precipitates in Al-Ag-Cu aged for 4\,h at 200\celsius.
a) and b) show bright field (BF) and HAADF STEM images from the same precipitate.
Distinct high and low contrast layers within the precipitate are visible in the bright field image. 
Viewing the image at a glancing angle shows that the matrix planes are in alignment on either side of the image indicating  that the matrix planes are commensurate across the precipitate. 
\label{fig-three-layer}}
\end{center}
\end{figure*}

High and low contrast layers were sometimes visible in images of thicker precipitates, 
suggesting that segregation of Ag and Al had taken place. 
Figure~\ref{fig-three-layer} shows a) bright field (BF) and b) HAADF STEM images of such a precipitate aged for 4\,h at 200\celsius.
In the enlarged insets in Figure \ref{fig-bf} it can be seen that the central region has some alternating contrast with three darker bands  separated from one another by a distance of $\sim$0.46\,nm i.e. close to the c-lattice parameter for \gammaprime. 
This layering extends over a thickness of seven close-packed planes, equivalent to three times the unit cell height  and the matrix planes are commensurate across the precipitate. 
Such commensurate precipitates were invariably found as isolated, single precipitates, rather than as members of the structured precipitate assemblies described in \cite{Rosalie2009a,Rosalie2009b}. 

\begin{figure*}
\begin{center}
\subfigure[One layer (Fig. \ref{fig-haadf-alagcu})]{\fig[0.4]{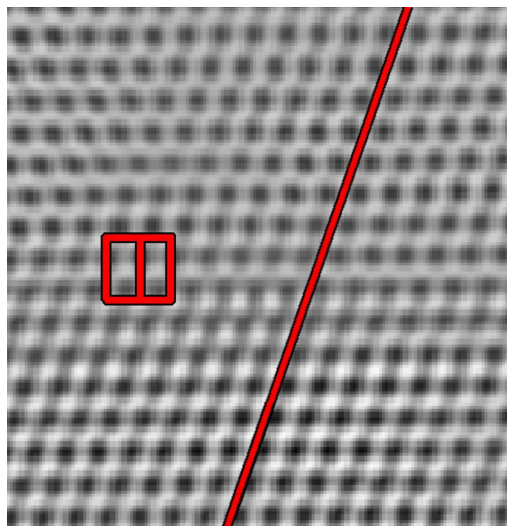}} \hfill 
\subfigure[Two layer (Fig.~\ref{fig-stem-4h})]{\fig[0.4]{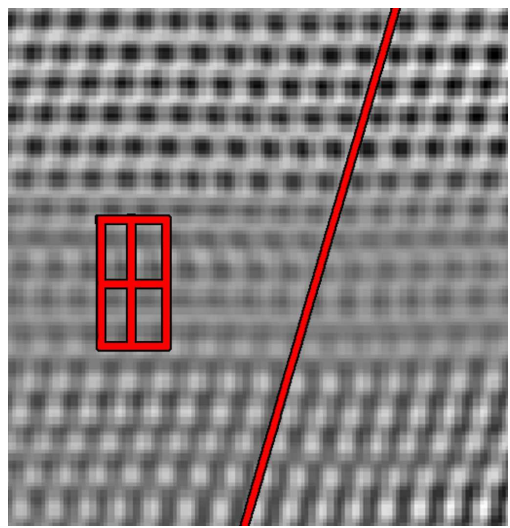}} 

\subfigure[Three layer (Fig.~\ref{fig-haadf-layers})]{\fig[0.4]{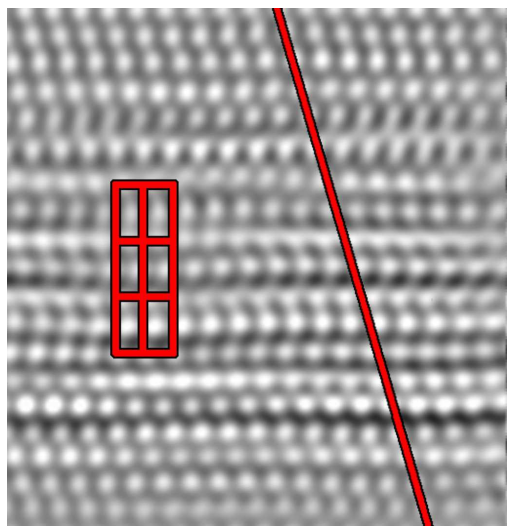}} \hfill 
\subfigure[FFT \label{Fft}]{\fig[0.24]{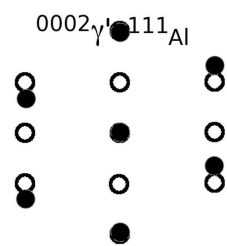}} \hfill 
\subfigure[Mask \label{mask}]{\fig[0.24]{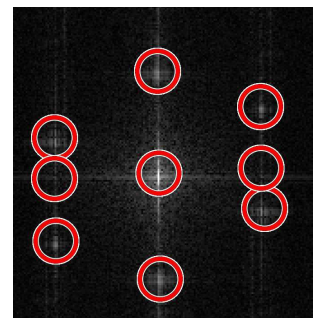}}  

\caption{Fourier-filtered STEM images.
The images showing  thickening of the hcp region  in steps of a single unit cell height. 
A sample Fourier transform and the locations of the reflections included in the inverse transform are shown in (d).
\label{IFFTs}}
\end{center}
\end{figure*}

The thickening behaviour can be most clearly seen if the STEM images are filtered by fast Fourier transform (FFT).
Figure~\ref{IFFTs} shows such the result of such filtering for the STEM  images in Figures \ref{fig-haadf-alagcu},  \ref{fig-stem-4h} and \ref{fig-haadf-layers}.
FFTs of the raw images were masked over first order spots for fcc and hcp structures and then inverse transformed. 
A representative Fourier transform is shown in Figure~\ref{Fft} together with a schematic of the reflections included in the inverse transform, (Figure~\ref{mask} Al; solid circles, \gammaprime; open circles). 
The masking excluded the central transmitted spot in order to remove atomic contrast and provide a clearer image of the structure.
A clear progression is visible in these images, with (a) showing a region with a single stacking fault in the fcc structure, suggesting the presence of a single transformation dislocation. 
The second image, (b) shows a region with two such changes in stacking occurring on alternate planes while in (c) three stacking faults are present, each on alternate planes.  
Overlays indicate the thickness of 1, 2 and 3 unit cell heights of the hcp region, respectively.  
Tracing the matrix planes across the precipitate shows that the 111 planes are incommensurate in (a) and (b) but and commensurate in (c) where 3 faults 
 are present.

\subsection{Segregation of silver}

\begin{figure*}
\begin{center}
\subfigure[Al-Ag 0.5\,h \label{fig-fwhm-b}]{\includegraphics[height=0.32\textwidth]{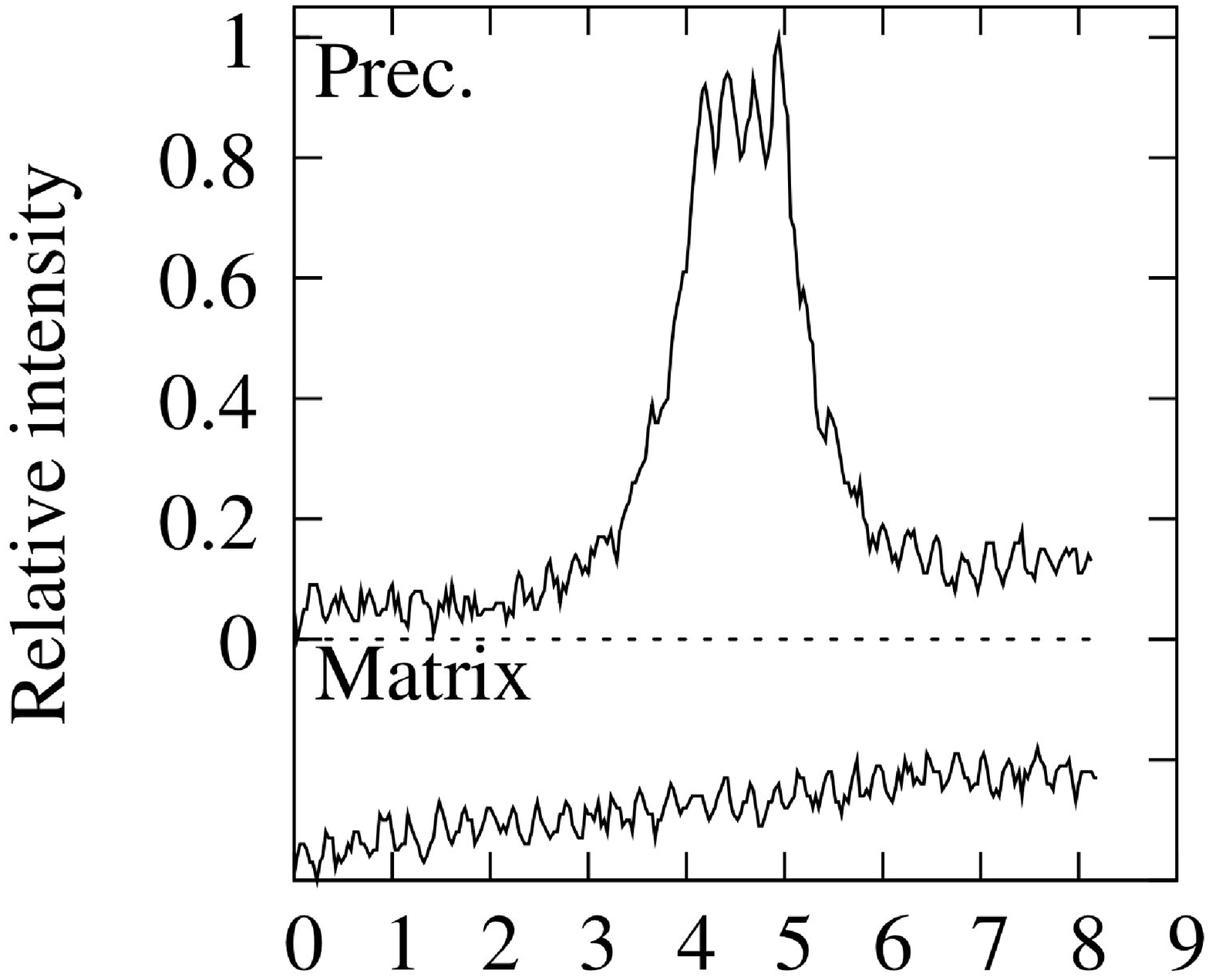}} \hfill 
\subfigure[Al-Ag-Cu 2\,h \label{fig-fwhm-c}]{\includegraphics[height=0.32\textwidth]{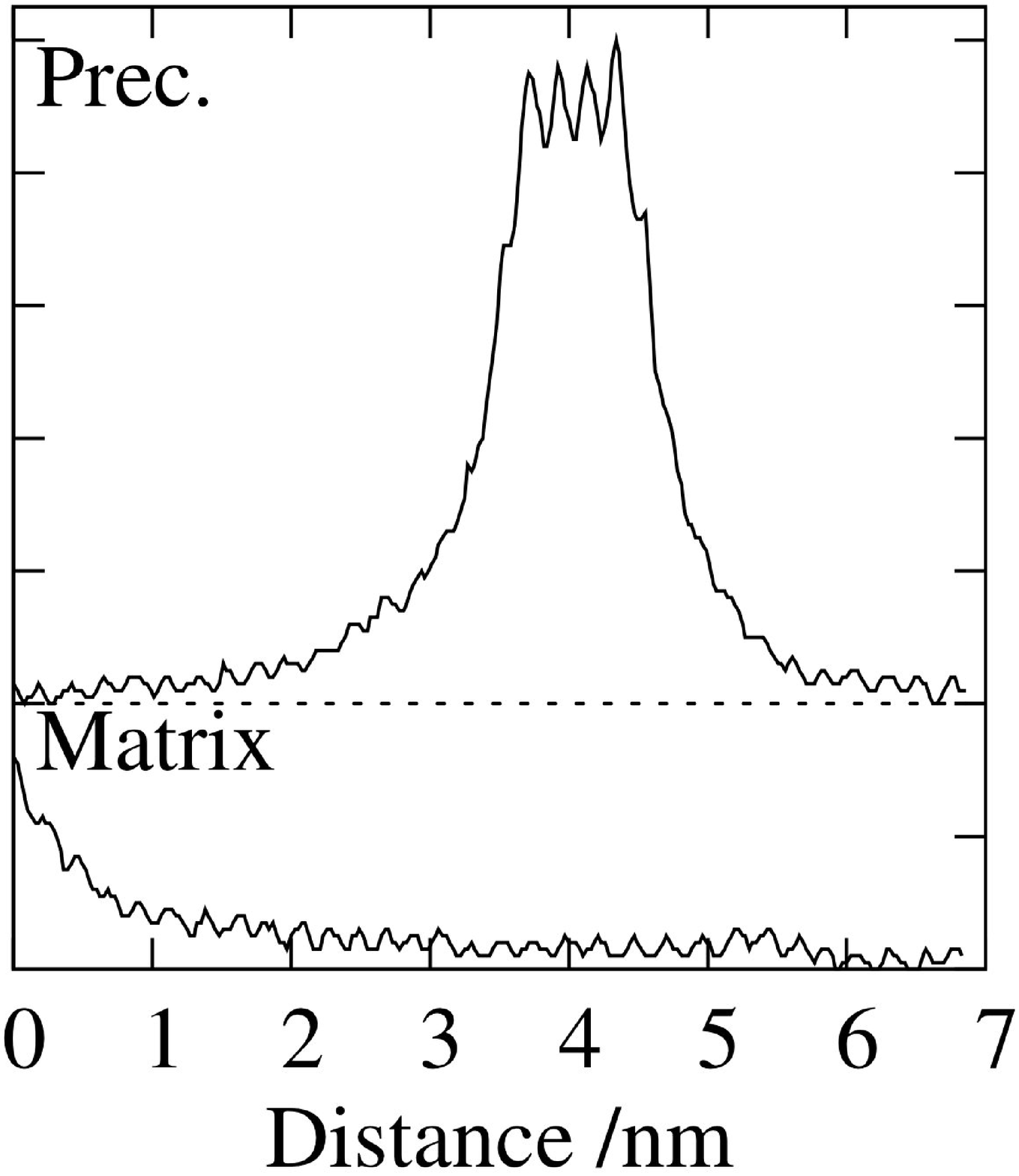}}\hfill 
\subfigure[Al-Ag-Cu h4\,h \label{fig-fwhm-d} ]{\includegraphics[height=0.32\textwidth]{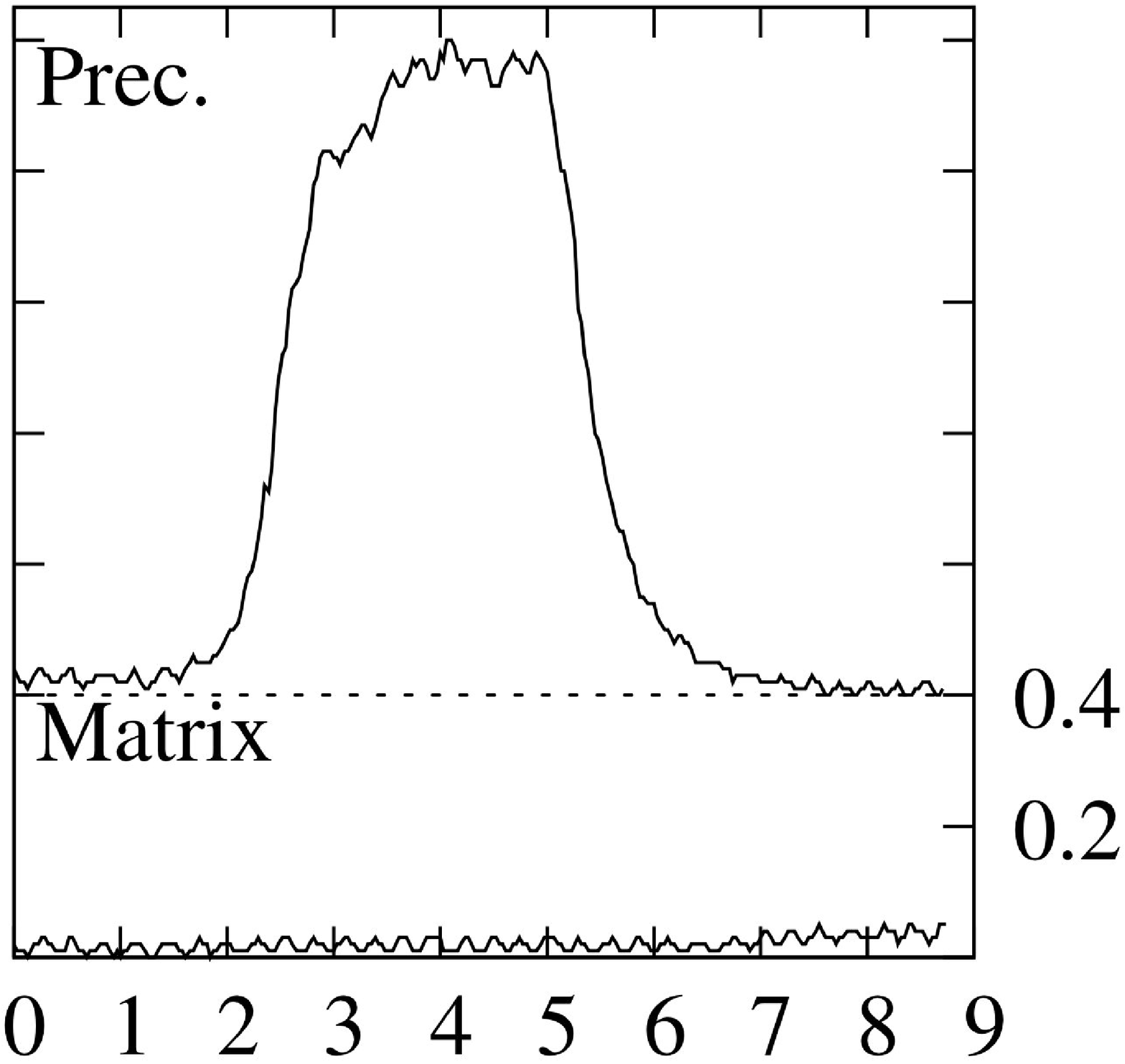}}
\caption{Line profiles of STEM HAADF intensity. 
a)-c) show representative HAADF STEM profiles obtained after 0.5\,h (Fig~\ref{fig-haadf-alag}), 2\,h and 4\,h (Fig~\ref{fig-stem-4h}) with Ag enrichment in a region considerably thicker than the hcp region.
\label{fig-fwhm}}
\end{center}
\end{figure*}

Atomic contrast images showed silver segregation extending well beyond the hcp region. 
Representative line profiles of the HAADF STEM intensity for ageing times of 0.5--4\,h are shown in Figure~\ref{fig-fwhm} and indicate significant segregation of silver to the defect/precipitate in the early stages of ageing.

The width of the silver-enrich region varied considerably  between precipitates  and there was a very gradual increase in the thickness with continued ageing. 
Precipitates examined after 0.5 ageing had a mean thickness 1.24\,nm
but correspond to those where close examination revealed that the change in stacking occurred across a single plane.  
There was no significant difference between the mean thickness value for Al-Ag (1.21\,nm)  and Al-Ag-Cu alloys (1.24\,nm). 
The precipitate thickness averaged 1.35\,nm after 2\,h of ageing. 
However, only after 4\,h of ageing did a few precipitates show a substantial increase in thickness, with  a few cases observed with thicknesses  
$\sim$2.8\,nm. 
Precipitates with thicknesses greater than 2.0\,nm corresponded to those (like those in Figures~\ref{fig-stem-4h} and  \ref{fig-three-layer}) where ordering was observed. 

\section{Discussion}

\subsection{The fcc(Al) to hcp(\gammaprime) phase transformation}

The finest scale structures observed in the Al-Ag(-Cu) alloys were silver-enriched single stacking faults.
The stacking fault was apparent in the displacement of the matrix planes in both HRTEM and HAADF STEM images and appeared to occur across a single close-packed plane. 
Precipitate diameters were as low as $4\pm0.5$\,nm (Figures~\ref{fig-hrtem-a} and \ref{fig-stem-a}). 
This is similar to the finest scale \gammaprime plates reported in cyclically-deformed Al-1\,at\%Ag \cite{voss:1999} and close  
to density functional theory predictions of the critical nuclei size \cite{Finkenstadt2010}. 
The scale of precipitation and the presence of a single fault both suggests that the  fault/precipitate structure should be close to that of the critical nuclei.

These structures are somewhat similar to the silver enriched single stacking faults proposed by Finkenstadt and Johnson \cite{finkenstadt:2006}.
However, ordering of silver atoms was not observed and instead silver was generally concentrated into 4 close-packed planes that displayed approximately equal intensity in HAADF STEM images.
The HAADF STEM intensity and silver concentration gradually diminished  in the surrounding atomic planes. 
In the absence of Al/Ag ordering into layers these structures should be more accurately described as silver-enriched faults, rather than a distinct phase. 

Ordering of the silver and aluminium atoms into distinct layers was often observed for  for triple-faulted structures, as in Figure~\ref{fig-bf-layers} and occasionally seen in precipitates containing two stacking faults. 
This suggests that ordering is a gradual process, rather than occurring simultaneously with the structural rearrangement. 
Since the \gammaprime structure is defined by both the the hcp crystal structure and the presence of alternating high and low silver layers,  it is argued that structures with two or more stacking faults and in which silver and aluminium have  begun to segregate are true \gammaprime precipitates, rather than silver-enriched stacking faults. 

It is not possible to completely rule out artefacts such as overlap with the matrix causing
such layered contrast. 
This appeared unlikely, however, given that the alternating contrast bands were only observed with a spacing of two close packed layers (as is expected for the \gammaprime phase)  and occur only in the region where there is also a change of stacking and not in the adjacent regions, which are also  silver-enriched.

The presence of structures with single, double and triple stacking faults can be explained if thickening of the hcp region can be explained by the passage of single (rather than paired) Shockley partial dislocations. 
Transformation dislocations were not directly observed; however, the change in stacking on alternate layers (with spacing $c(\gammaprime) =0.46$\,nm) and the formation of commensurate precipitates for three changes in stacking is consistent with models for the Al(fcc)$\rightarrow$\gammaprime phase transformation \cite{muddle:1994} and   
the formation of strain-accommodated hcp precipitates \cite{shchegoleva:1976}. 

It should be noted that equal numbers of each of the three variants of Shockley partial dislocations in a given \{111\} plane are necessary to balance the shape strain. 
In this case  the individual displacements cancel and the net matrix displacements will be zero. 
A double-faulted structure with height = 2$\times c$ \gammaprime is not able to self-accommodate the shape strain.
However, apparently self-accommodated  precipitates (as in Figures~\ref{fig-three-layer} and   \ref{IFFTs}) were uncommon, suggesting that thickening is sluggish under these conditions. 

Based on the present observations it is proposed that \gammaprime precipitation commences with segegration to stacking faults as proposed by Finkenstadt and Johnson \cite{finkenstadt:2006} and then proceeds (in the initial stages at least) via the progression of single ledges with Shockley partial dislocation character.

\begin{figure*}
\begin{center}
\subfigure[\label{fig-nucleation-a}Single fault]{\fig[0.32]{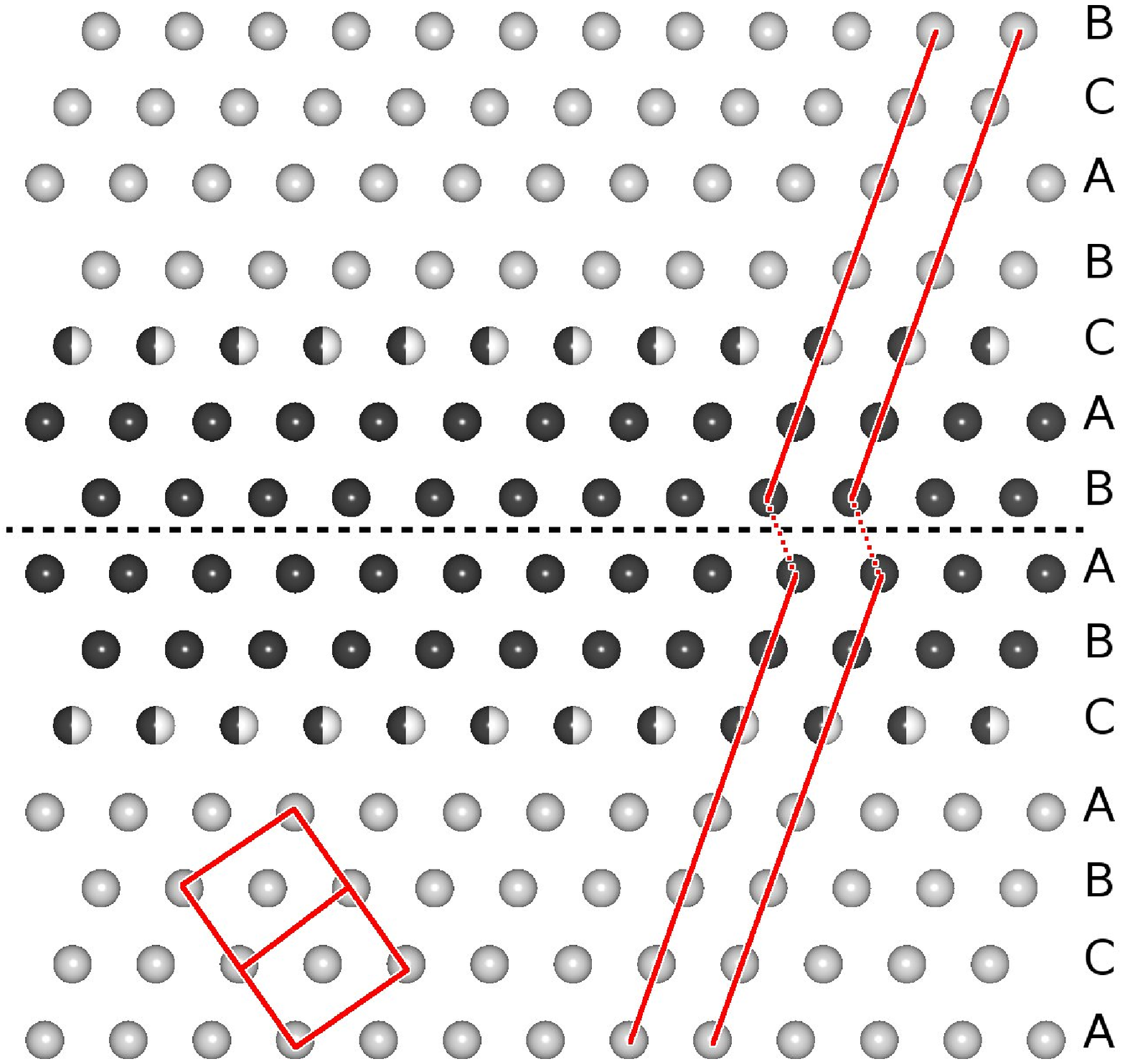}} 
\subfigure[\label{fig-nucleation-b}Double fault]{\fig[0.32]{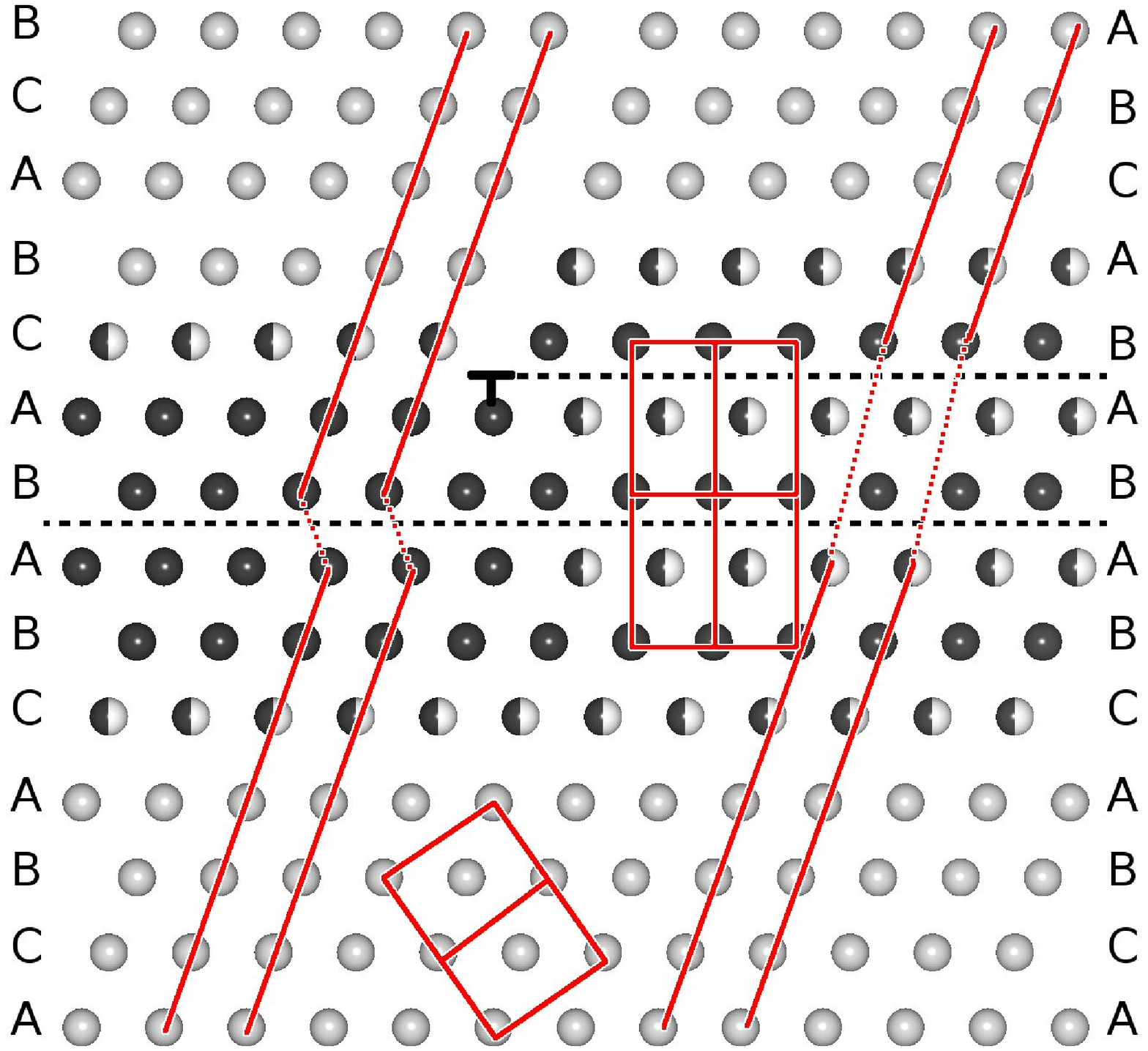}} 
\subfigure[Triple fault\label{fig-nucleation-c}]{\fig[0.32]{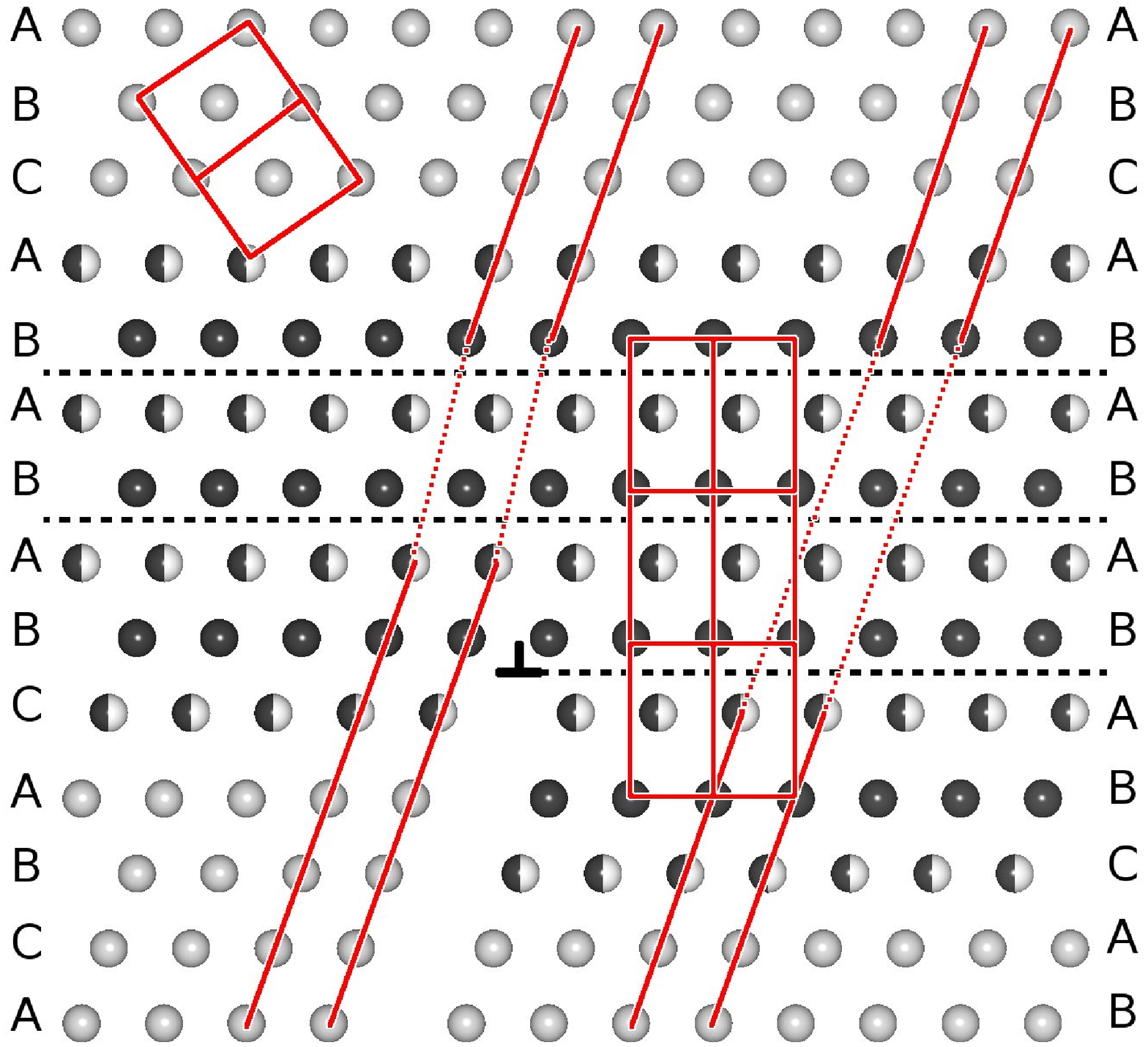}} 
\caption[A model for the nucleation of  \gammaprime]{A model for the nucleation of  \gammaprime.  
(a) Ag-enriched  stacking fault with central four layers heavily enriched in silver, with lower levels of silver in the adjacent layers. 
 b) The structure during the passage of a Shockley partial dislocation along the upper surface of the precipitate, 
leaving a second stacking fault.
To the right of the dislocation core silver and aluminium atom have begun to order, forming the layered  \gammaprime structure. 
The matrix planes remain out of alignment in this condition. 
c) illustrates the passage of an additional Shockley partial dislocation to form a  third unit cells height. 
\label{fig-nucleation}}
\end{center}
\end{figure*}

This mechanism is presented schematically  in Figure~\ref{fig-nucleation} with Al and Ag atoms indicated by light and dark circles, respectively. 
The initial stacking sequence is presented to the right of the figure with the stacking fault indicated by a dashed line.
Lines drawn along matrix \planes[]{111} planes are displaced in the habit plane on passage planes through the silver-enriched fault, with the displacement equivalent to that of a Shockley dislocation in the habit plane. 

Figure~\ref{fig-nucleation-a} shows a four heavily silver-enriched close-packed layers surrounding a stacking fault. 
Such structures were observed in HAADF STEM images, with silver enrichment across up to 8 close-packed planes (as in Figure~\ref{fig-haadf-alagcu}) 
\footnote{For clarity only a single layer additional layer surrounding the defect is shown as silver-rich}.

The successive passage of Shockley partial dislocations in Figure~\ref{fig-nucleation-b} and (c) adds additional stacking faults and increase the thickness of the hcp zone by 2 close-packed layers ($c (\gammaprime))$ each time. 
(To show the stacking sequence more clearly, the atomic positions have not been relaxed.)
The stacking sequence before and after passage of the defect are shown to the left and right of the diagram respectively.

In the double faulted structure the matrix planes are displaced with respect to one another on either side of the precipitate. 
This proposed structure might be seen to correspond to the  5-layer precipitates in which there was a displacement of the matrix planes 
(Figure~\ref{fig-stem-4h}).
Precipitate plates of similar thickness were reported in cylically-deformed Al-Ag where it was noted that only two variants of the Shockley partial dislocation were present \cite{voss:1999}. 
While it was argued that this could reflect the formation of the precipitate on dislocations the observations would also be compatible with non self-accommodated \gammaprime plates of $2\times c_{\gammaprime}$ thickness.

The nucleation of a second transformation dislocation in Figure~\ref{fig-nucleation-c}  would increase the thickness of the hcp region further, and returns the matrix planes to full alignment. 
This is now structurally identical to the arrangement proposed by Shchegoleva \cite{shchegoleva:1976}, but 
with nucleation having occurred on a pre-existing stacking fault rather than homogeneously. 

\subsection{The barrier to nucleation of the \gammaprime (\gammachem) phase}
The activation energy barrier to nucleation is classically regarded as being composed of interfacial and strain energy terms. 
Recent density functional theory calculations suggest that the coherent $\gammaprime_{0001}/\alpha_{111}$ interface has a relatively low energy of $\sim 15\pm10 \mathrm{mJ/m^2}$  \cite{Finkenstadt2010}.
The volumetric strain generated by the \gammaprime phase is negligible, as precipitation results in a relative dilational volumetric strain $\Delta V$ of -0.0255  \cite{nie:1999}. 
This relatively low energy barrier is difficult to reconcile with poor nucleation behaviour of the \gammaprime phase. 

The observations in the present work are consistent with a precipitate which is difficult to nucleate. 
The majority of structures, for example, still contained only a single stacking fault after 4\,h ageing.
 In addition, HAADF STEM images showed the silver-enriched region extended well beyond the faulted zone, with a mean thickness of 1.24\,nm,  $(2.7\times c(\gammaprime))$
despite the presence of only a single stacking fault (as in Figures~\ref{fig-stem-4h} and \ref{fig-fwhm}).
This is a strong indication that silver segregation is occurring in advance of the fcc $\rightarrow$ hcp transformation 
 (e.g. as suggested by Figure~\ref{fig-haadf-alagcu}) and that the transformation is sluggish despite abundant silver around the defect. 

The nucleation behaviour of \gammaprime could be explained if the transformation had characteristics of both diffusional and displacive phase transformations \cite{muddle:1994a, christian:1997}, requiring the change in the shape of the transformed volume to be considered.   
This approach would involve a latttice correspondence between matrix and precipitate like those in classical displacive transformations \cite{wechsler:1953,bowles:1954a, bowles:1954b}. 
Such lattice correspondences can be defined for diffusional or ordering transformations (e.g. Au-Cu, \cite{smith:1960}),  
where the  product and matrix share a low energy planar coherent (or semi-coherent) interface \cite{muddle:1994}.
The shear strain frequently far exceeds the volumetric strain \cite{nie:1999} and for \gammaprime nucleation  
the lattice correspondence  predicts a shear strain of 0.354 compared to a volumetric strain of -0.0255  \cite{muddle:1994a}.  
Traditionally, however, the displacive and diffusion mechanisms of phase transformations  are commonly regarded as mutually exclusive \cite{aaronson:1972, aaronson:1990},
with long-range diffusion during diffusional transformations (like \gammaprime nucleation) essentially reconstructing the transformed volume \cite{muddle:1994a}.

It has also been proposed that such shear strains could be neutralised via the formation of self-accommodating structural units, thus negating any local shape change in the transformed volume\cite{howe:1985a}. 
Such units would necessarily have a minimum height of $3\times c(\gammaprime)$ and the present observations of 1,2 and 3-faulted structures suggest that this is not the case during the earliest stages of ageing. 

The shear strain for non-accommodated \gammaprime plates can be estimated using the analysis of Eshelby, as set out by Christian \cite{christian:1958}. 
The influence of the shear strain drops off rapidly with increasing precipitate diameter; 
however, assuming an interfacial energy of 15\,$\mathrm{mJ/m^2}$ \cite{Finkenstadt2010} then the shear strain energy is equal to the interfacial energy for precipitates of radius $\sim$7\,nm. 
Precipitates of similar or lower diameter have been reported in Al-Ag \cite{voss:1999} and Al-Ag-Cu alloys \cite[Fig 5b]{Rosalie2009b} and were also seen in the present investigation. 
This suggesting that it is feasible for shear strain to contribute significantly to the overall barrier to nucleation for precipitates of this scale.  

\section*{Conclusions}

The nucleation and early growth of the \gammaprime (\gammachem) precipitate has been examined in detail.
This process commenced with the segregation of silver to stacking faults associated with dislocation loops in the aluminium alloy. 
The absence of Ag-Al ordering indicates that these structures are not genuine \gammaprime precipitates, but represent a pre-precipitat\-ion stage.  
Thickening of the hcp region involved the generation of additional stacking faults,  bounded by Shockley partial transformation dislocations to give double or tripled faulted regions. 
This process, combined with gradual ordering of Ag and Al  lead to the formation of the \gammaprime phase precipitate. 
Thickening of the precipitates was sluggish, despite extensive silver segregation  around the stacking-fault. 
The double unit-cell structure was not able to self-accommodate the shape strain associated with the precipitate, a factor which 
would required the inclusion of an additional shear strain energy term in the activation energy barrier to nucleation.
The shear strain energy for precipitates of this scale is of similar magnitude to the interfacial energy and its inclusion in the total energy barrier may explain the poor nucleation potential of the \gammaprime phase.

\section*{Acknowledgements}
The authors gratefully acknowledge the support of the Australian Research Council through the Centre of Excellence for Design in Light Metals.
One of the authors (JMR) also gratefully acknowledges the support of the Japan Society for the Promotion of Science (JSPS) through a JSPS fellowship. 
The authors acknowledge use of the facilities at the Monash Centre for Electron Microscopy and engineering support by Russell King.

\end{document}